\documentclass[aip,jcp,amsmath,reprint]{revtex4-1}

\usepackage{graphicx}
\usepackage{hyperref}
\usepackage[capitalize]{cleveref}
\crefname{section}{Sec.}{Sec.}
\crefname{figure}{Fig.}{Fig.}
\renewcommand{\vec}[1]{\mathbf{#1}}

\usepackage{color}

\begin{document}

\title{Efficient Equilibration of Hard Spheres with Newtonian Event Chains}
\author{Marco Klement}
\email{marco.klement@fau.de}
	
\author{Michael Engel}
\email{michael.engel@fau.de}

\affiliation{Institute for Multiscale Simulation, Friedrich-Alexander University Erlangen-N{\"u}rnberg, 91058 Erlangen, Germany}

\date{\today}

\begin{abstract}
An important task in the simulation of hard spheres and other hard particles is structure prediction via equilibration.
Event-driven molecular dynamics is efficient because its Newtonian dynamics equilibrates fluctuations with the speed of sound.
Monte Carlo simulation is efficient if performed with correlated position updates in event chains.
Here, we combine the core concepts of molecular dynamics and event chains into a new algorithm involving Newtonian event chains.
Measurements of the diffusion coefficient, nucleation rate, and melting speed demonstrate that Newtonian event chains outperform other algorithms.
Newtonian event chains scale well to large systems and can be extended to anisotropic hard particles without approximations.
\end{abstract}

\maketitle


\section{Introduction}

Hard particles receive much attention as model particles in the statistical mechanical theory of colloidal self-assembly.\cite{Frenkel2015,Manoharan2015,Boles2016}
Research is often tasked with finding the stable phase, which can be performed in simulation via equilibration from a disordered fluid.\cite{Agarwal2011,Damasceno2012a,Ni2012}
During equilibration, the system follows a trajectory in configuration space from a basin of low probability corresponding to a phase with higher free energy into a basin of high probability corresponding to a phase with lower free energy.
Configuration space can either be sampled deterministically with molecular dynamics or stochastically with Monte Carlo.
The equilibrium phase, reached after a sufficiently long simulation, is independent of the algorithm used.
However, the time to reach equilibrium depends crucially on details of the simulation trajectory.
Many ideas have been proposed to improve efficiency by decreasing the length of the trajectory connecting a starting configuration to the basin of lowest free energy.

A classic way to simulate particles is to solve Newton's equations of motion.
This is the approach of molecular dynamics.
In the case of hard particles, molecular dynamics involves the prediction of collision events.\cite{Alder1959}
Event-driven molecular dynamics (EDMD) moves all particles fully collectively and thus equilibrates density fluctuations with the speed of sound.
EDMD is generally considered to be highly efficient.\cite{Isobe1999,Donev2005a,Engel2013,Isobe2016}
In contrast, local Monte Carlo (LMC) moves only one randomly chosen particle at a time.\cite{Metropolis1953}
LMC neglects particle momentum, which means fluctuations propagate and relax much slower resulting in poor performance in comparison to EDMD.
Advantages of LMC are its simple implementation, good scaling to large systems, and easy generalization to anisotropic particles.
Equilibration via Monte Carlo that hopes to rival EDMD in simulation efficiency requires collective moves.~\cite{Dress1995,Liu2004,Whitelam2007}
A promising approach is updating particles in chains.\cite{Jaster1999}
Chain moves can be performed rejection-free in the form of event chain Monte Carlo.\cite{Bernard2009,Peters2012,Michel2014,Kampmann2015,Isobe2015,Harland2017,Weigel2018}
In an event chain, a randomly selected particle is displaced until it collides with another particle.
Next, the collision partner is displaced until it collides with yet another particle and the process iterated.
The chain of collision events terminates once a preselected chain length has been reached.\cite{Bernard2009}

Existing collective Monte Carlo algorithms neglect a core concept of molecular dynamics, the momentum of the particles in the system.
As a result, fluctuations decay slowly, which lowers efficiency.
Here, we propose combining the main advantage of EDMD, Newtonian dynamics, with the main advantage of event chain Monte Carlo, collective moves in chains.
We call the new algorithm Newtonian event chains (NEC).
We evaluate equilibration efficiency of NEC and compare it to existing algorithms.
While EDMD is most efficient per particle displacement, NEC comes close and is more efficient when analyzed in terms of CPU time.
The only free parameter is the duration of each event chain.
In the limit of short chains, NEC reduces to a variant of molecular dynamics with random permutation of particle updates.


\section{Algorithms and methods}

\subsection{Existing hard sphere simulation algorithms}

\begin{figure*}
\includegraphics[width=\textwidth]{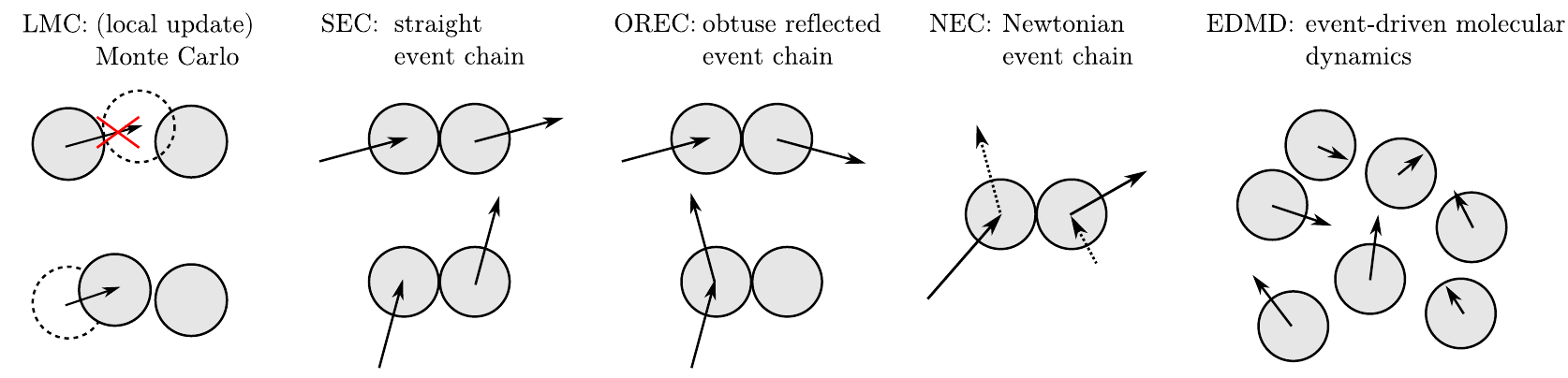}
\caption{Schematic overview of simulation algorithms for hard spheres. 
LMC accepts moves that do not generate overlaps.
A single rejected (top) and accepted (bottom) trial move is shown.
SEC, OREC, and NEC consist of chains of collision events.
A single collision event is shown.
SEC and OREC distinguish acute collision angles (top) and obtuse collision angles (bottom).
NEC updates both particle velocities at the collision.
Finally, EDMD moves all particles fully collectively according to Newtonian dynamics.
}
\label{fig.chainmodes}
\end{figure*}

\emph{Local Monte Carlo (LMC).}
A local Monte Carlo trial move consists of selecting a random particle and displacing it to a new position randomly chosen within a sphere of radius $L_\text{step}$.
The trial move is accepted with probability given by the Metropolis criterion.
For hard particles the Metropolis criterion states that moves that generate one or more overlaps are always rejected (\cref{fig.chainmodes}).
Trial moves that do not generate an overlap are always accepted.
LMC is simple to implement but not efficient.

\emph{Event-driven molecular dynamics (EDMD).}
In molecular dynamics each particle has position and velocity, which are updated by solving Newton's equations of motion.
Molecular dynamics for hard particles can only be performed event-driven.
This means the times of all collision events in the system have to be computed.
Collisions are sorted and evaluated in temporal order, which is slow and difficult to scale to large systems.
The pair of particles involved in the current collision event is moved up to their collision positions.
Next, new velocities are computed from the equations for elastic collisions.
EDMD is difficult to parallelize and scales poorly to large systems.
It also cannot be applied to anisotropic particles without an iterative solver, which lowers precision and performance, because calculating collision times for such particles involves nonlinear equations that usually have no analytic solution.\cite{Donev2005a}

\emph{Event chain Monte Carlo.}
An event chain starts with a randomly selected particle $i$ with radius $r_i$, located at position $\vec{x}_i$ and a normalized displacement vector $\vec{d}$.
Particle $i$ is displaced along $\vec{d}$ until it collides with a first other particle $j$, its collision partner.
The position of particle $i$ is then updated to be the contact point with the collision partner, given by
\begin{equation}
\vec{x}_i' = \vec{x}_i + \vec{d}\left(\vec{d} \cdot \vec{x}_{ij} - \sqrt{ (r_i+r_j)^2 - x_{ij}^2 + (\vec{d} \cdot \vec{x}_{ij})^2 }\right)
\end{equation}
with $\vec{x}_{ij}=\vec{x}_j - \vec{x}_i$.
Next, it has to be decided with which particle and in which direction the event chain is continued.
We denote the new particle that is chosen to continue the event chain as $i'$ and the new displacement direction with $\vec{d}'$.

Several variants for selecting $i'$ and $\vec{d}'$ have been proposed in the literature.\cite{Bernard2009,Weigel2018}
Straight event chains (SEC) is the simplest choice.
The chain is continued with the collision partner and in the same direction,
\begin{equation}
\text{SEC:}\quad i'=j,\quad\vec{d}' = \vec{d}.
\end{equation}
In contrast, in reflected event chains (REC) the direction is reflected,
\begin{equation}
\text{REC:}\quad i'=j,\quad\vec{d}' = \vec{d}_r = 2 \vec{x}_{ij} \frac{\vec{d}\cdot\vec{x}_{ij}}{x_{ij}^2} - \vec{d} \label{eq-ec-REC}.
\end{equation}
Detailed balance is obeyed in REC because the equation for $\vec{d}'$ is symmetric.
The event chain ends when it reaches a preselected length.
Once the sum of all displacements distances $\|\vec{x}_i'-\vec{x}_i\|$ adds to above the chain length $L_\text{chain}$, the last particle is displaced only by the remaining distance.

We modify the event chain variants SEC and REC to mimic trajectories generated by Newtonian dynamics more closely.
For this purpose an event chain is sought that follows the path of largest momentum transfer.
This is achieved by distinguishing two cases.
Collisions with obtuse collision angles, $\vec{d}\cdot\vec{d}_r>0$, are handled as before in REC.
Collisions with acute angles, $\vec{d}\cdot\vec{d}_r<0$, are now handled as total reflections.
Total reflections continue the chain with the original particle and mirror the displacement vector.
We call the resulting algorithm obtuse reflected event chains (OREC).
It obeys
\begin{equation}
\text{OREC:}\quad
\begin{cases}
i'=j,\quad\vec{d}' = \vec{d}_r  & \text{if }\vec{d}\cdot\vec{d}_r>0,
 \\
i'=i,\quad\vec{d}' = -\vec{d}_r & \text{if }\vec{d}\cdot\vec{d}_r<0.
\end{cases}
\label{eq-ec-decision}
\end{equation}
Again, the equation for $\vec{d}'$ is symmetric, which ensures detailed balance.
Other variants of event chains handling collisions with acute and obtuse angles in different ways are possible.
We tested other variants but found that OREC is the most efficient variant among those.
For this reason we will focus on the well-known event chain variant SEC and its slight modification OREC in the following.


\subsection{Newtonian event chains}

Newtonian dynamics is beneficial for efficient equilibration.
We therefore wish to augment event chains by assigning a velocity vector $\vec{v}_i$ to each particle in the system.
Velocities are initialized according to the Maxwell-Boltzmann distribution.
As before, we start an event chain by selecting a particle $i$ at random.
This particle is then displaced along its normalized velocity vector $\vec{d}=\vec{v}_i/v_i$ up to the first collision with collision partner $j$.
The collision is handled simply as an elastic collision of two spheres with the same mass taking into account the velocity vectors of both particles.
From the collision equations we obtain the new velocities $\vec{v}'_i$ and $\vec{v}'_j$.
Next, the chain proceeds with the collision partner as in SEC but now in direction of the updated velocity.

The NEC algorithm requires only a few more calculations than the SEC algorithm but otherwise no additional effort.
It reads
\begin{align}
\text{NEC:}\quad&\vec{v}_i' = \vec{v}_i + \vec{x}_{ij}\frac{\vec{v}_{ij}\cdot\vec{x}_{ij}}{x_{ij}^2},\\
&\vec{v}_j' = \vec{v}_j - \vec{x}_{ij}\frac{\vec{v}_{ij}\cdot\vec{x}_{ij}}{x_{ij}^2},\\
& i'=j,\quad\vec{d}' = \frac{\vec{v}'_j}{v'_j}
\end{align}
with $\vec{v}_{ij}=\vec{v}_j - \vec{v}_i$.
The event chain ends when it reaches a preselected duration.
Once the sum of all displacement times $\|\vec{x}_i'-\vec{x}_i\| / v_i$ adds to above the chain duration $T_\text{chain}$, the last particle is displaced only by the remaining time.

It is important to preselect a duration for event chains in NEC as parameter instead of preselect a length.
To see this, assume that all particle velocities are initially distributed according to Maxwell-Boltzmann statistics.
Because we perform collisions using Newtonian dynamics, the velocity distribution after a fixed elapsed time or, alternatively, a preselected event chain duration is not affected.
As demonstrated in Ref.\citenum{Manousiouthakis1999}, the invariance of the Maxwell-Boltzmann velocity distribution guarantees local detailed balance, which in turn is sufficient to ensure a statistically valid Monte Carlo simulation.
Hence, NEC with preselected chain duration is a statistically correct algorithm.

In contrast, let us assume a preselected length was chosen for event chains in NEC.
Internal time of event chains that contain many slow particles then advances faster than internal time of event chains that contain many fast particles.
As a result, such event chains are slightly more likely to terminate with a fast particle than expected from Maxwell-Boltzmann.
This means the velocity distribution skews towards higher velocities.
Clearly, NEC with a preselected chain length violates the local balance condition.
We directly observe this effect.
Reference values for pressure and local order are reproduced correctly when the chain duration is fixed (\cref{secsec.validation}) but not if the chain length is fixed.


\subsection{Measured quantities} \label{sec.measuredQuantities}

There are many ways to evaluate equilibration efficiency, including autocorrelation times of thermodynamic quantities,\cite{swendsen1987nonuniversal,Jaster1999,Weigel2018} positional auto correlation such as the diffusion coefficient,\cite{Bernard2009,Ruzicka2014,Isobe2015} and evolution of specific quantities.\cite{Whitelam2007,Isobe2015}
Here we focus on measuring the diffusion coefficient, nucleation rate, and melting speed.
We find that these three quantities are roughly equivalent for evaluating the efficiency of hard sphere simulation algorithms.
Because Monte Carlo does not have an internal time, we analyze the advancement of the simulation by the number of displacements $N_\text{disp}$.
This number is the number of trial moves for LMC, the number of particle translations in all event chains, and the number of time steps multiplied with the number of particles $N$ in the system for EDMD.
We interpret $N_\text{disp}$ as a measure for the length of the trajectory in configuration space.

\emph{Diffusion coefficient.}
The diffusion coefficient as a function of the number of displacements is calculated from the mean square displacement by the relationship
\begin{equation}
D_\text{disp} = \lim_{N_\text{disp}\rightarrow\infty}\frac{\sum_{i=1}^{N} \left(\Delta\vec{x}_i - \langle\Delta\vec{x}\rangle\right)^2}{6 \; N_\text{disp}},
\end{equation}
where $\Delta\vec{x}_i$ is the total displacement vector of particle $i$ after $N_\text{disp}$ particle displacements in the system and we subtract the mean displacement vector $\langle\Delta\vec{x}\rangle$ obtained by averaging over all particles.
Not all displacements require the same computational cost.
We therefore also determine the diffusion coefficient as a function of CPU time $t_\text{cpu}$,
\begin{equation}
D_\text{cpu} = \lim_{t_\text{cpu}\rightarrow\infty}\frac{\sum_{i=1}^{N} (\Delta\vec{x}_i - \langle\Delta\vec{x}\rangle)^2}{6\;N\; t_\text{cpu}}.
\end{equation}
While $D_\text{disp}$ is an absolute measure of the effectiveness of displacements for generating diffusion, $D_\text{cpu}$ is affected by practical factors such as the implementation of the simulation algorithm and the computer hardware.

\emph{Nucleation rate.}
We model nucleation as a Bernoulli process.
The probability to find a new nucleus within time $\tau$ is $p$.
There are $N_t=t/\tau$ trials to create a nucleus within the simulation time $t$.
On average we find $N_n=N_t p$ nuclei with a variance of $\text{Var}(N_n) = N_t p (1-p)$.
For sufficiently small $\tau$ we can approximate $1-p \approx 1$, and thus $\text{Var}(N_n) = N_t p = N_n$.
The nucleation rate $R = N_n / t$ has standard deviation $\sigma = \sqrt{N_n}/t$.
Because Monte Carlo does not have an internal time, we use the number of displacements $N_\text{disp}$ or CPU time $t_\text{cpu}$ instead.
Nucleation rate is measured by performing several independent simulations starting in a fluid phase.
When we detect a critical nucleus, we increment $N_n$ and reset the simulation to the fluid phase.
We stop once we find the twentieth nucleus and record $t_\text{cpu}$ and $N_\text{disp}$ at the end of the sequence.

\emph{Structure analysis.}
We quantify the ordering process during the fluid-to-solid transformation with the global bond orientational order parameter $Q_6$ and the local bond orientational order parameter $q_6$.\cite{Steinhardt1983}
The SANN algorithm\cite{Meel2012} is used to identify nearest neighbors.
Nuclei are detected with the common cluster algorithm.\cite{Filion2010a}
We determine melting speed by measuring the necessary CPU time to reach $Q_6=0.1$.


\subsection{Implementation and setup}

\begin{table}
	\setlength{\tabcolsep}{3mm}
	\begin{tabular}{l c c c}
		\hline
		\hline
		& \multicolumn{3}{c}{$N_\text{disp}$/h of CPU time} \\
		& \multicolumn{2}{c}{this work, 3.5~GHz} & Ref.\citenum{Isobe2015}, 2.8~GHz\\
		& $N=2^{14}$ & $N=2^{17}$ & $N=2^{17}$ \\
		\hline
		LMC  & $7.8\times10^9$ &  $8.6\times10^9$ &  $6.5\times10^9$ \\
		SEC  & $3.2\times10^9$ &  $2.5\times10^9$ &  $3.15\times10^9$ \\
		OREC & $3.2\times10^9$ &  $2.8\times10^9$ &  -- \\
		NEC  & $3.1\times10^9$ &  $2.8\times10^9$ &  -- \\
		EDMD & $1.5\times10^9$ & $0.83\times10^9$ & $0.46\times10^9$ \\
		\hline
		\hline
	\end{tabular}
	\caption{
		Computational costs for displacing particles for two system sizes and comparison to literature.
		We list the number of displacements $N_\text{disp}$ per hour of CPU time.
		$N_\text{disp}$ corresponds to the number of trial moves for LMC, the number of particle translations for the event chain algorithms SEC, OREC, NEC, and the number of particle collisions for EDMD.		
		Data is obtained for an over-critical fluid at 54.8\% volume fraction prior to nucleation.
	}
	\label{tab.speedcomp}
\end{table}

We use the hard particle Monte Carlo (HPMC) package\cite{Anderson2015a} of HOOMD-blue\cite{Anderson2008a,Glaser2015} without changes as implementation for LMC.
All variants of event chain algorithms are implemented without parallelization into a private fork of HPMC.
Implementations employ an AABB tree as neighbor list.
EDMD is our own implementation using a priority queue with constant complexity as described in Refs.~\citenum{Wang2018,Bommineni2018}.
All simulations are performed on Intel Xeon E3-1240 v5 CPUs with 3.5~GHz.

We compare the computational costs for displacing particles in \cref{tab.speedcomp}.
LMC is the simplest algorithm and displaces particles the fastest.
The three event chain algorithms displace particles roughly in the same time, which is slower by a factor of 2.5 to 3.5 compared to LMC, depending on whether the system size is $N=2^{14}=16\,384$ or $N=2^{17}=131\,072$.
EDMD is the algorithm that requires most computational effort.
It displaces particles a factor of between 5 and 14 slower than LMC.
The performance of our implementations is comparable to that of Ref.~\citenum{Isobe2015}, which has been obtained, however, using Intel Xeon E5-2680 CPUs with 2.8~GHz, i.e.\ a similar microarchitecture with 20\% slower clock rate than the CPUs used in our tests.
If computational cost is corrected via division by CPU clock rate, which certainly is a strong simplification but nevertheless provides a rough comparison, then our implementation is 6\% faster for LMC, 37\% slower for SEC, and 44\% faster for EDMD.
An explanation for the slower performance of SEC is discussed at the end of \cref{Subsec.melt}.

Systems of $N$ spheres with diameter $d$ are initialized in either a random fluid starting configuration or a face-centered cubic (FCC) solid.
Simulations are performed in the isochoric ensemble.
Pressure is calculated from the virial expression as in previous works.\cite{Engel2013,Michel2014,Isobe2015}
Except for the parameter study optimizing Monte Carlo parameters in the next section, the step size of LMC is tuned for maximal speed to about 15\% to 20\% trial move acceptance probability.
Chains with chain length of approximately one particle diameter or equivalent chain duration are used for all event chain variants except when noted otherwise.


\begin{figure}
\includegraphics[width=\columnwidth]{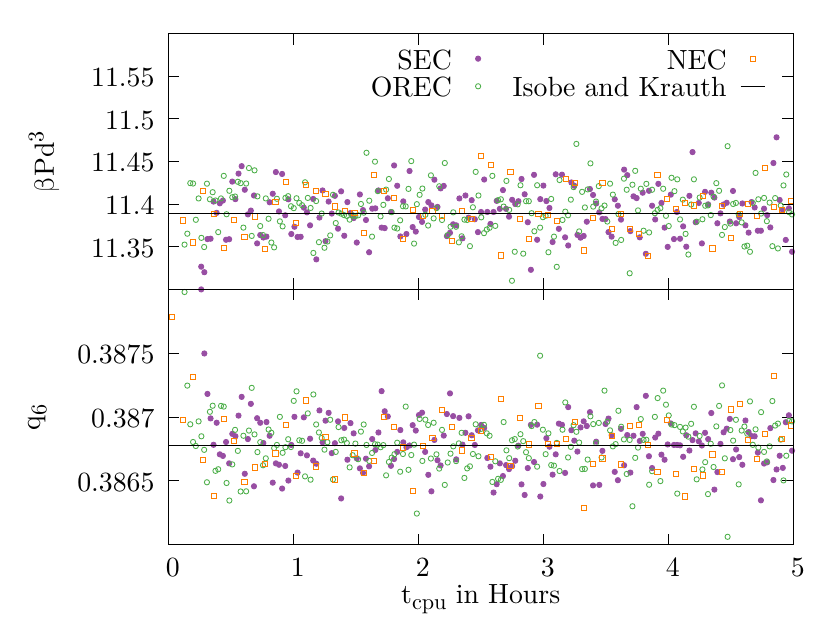}
\caption{
Dimensionless pressure $\beta P d^3$ and local bond orientational order parameter $q_6$ of hard spheres at $49\%$ volume fraction averaged over five hour simulations.
We compare the three event chain variants SEC, OREC, and NEC to literature data in \citet{Isobe2015}.
}
\label{fig.validation}
\end{figure}

\subsection{Validation}\label{secsec.validation}
To validate our implementation of the three event chain variants, we measure dimensionless pressure and the local bond orientational order parameter in a fluid of $N=131\,072$ particles at $49\%$ volume fraction and compare both to the reference values\cite{Isobe2015} $\beta P d^3 = 11.3894(1)$ and $q_6=0.38678$.
All our algorithms give results in agreement to these numbers within their respective standard errors (\cref{fig.validation}).
Pressure averages to $\beta P d^3 = 11.3907(14)$ for SEC, $\beta P d^3 = 11.3906(13)$ for OREC, and $\beta P d^3 = 11.3886(18)$ for NEC.
The equilibrium value for the local bond orientational order parameter is $q_6 = 0.38678(3)$ for SEC, $q_6 = 0.38679(3)$ for OREC, and $q_6 = 0.38680(4)$ for NEC.
Our standard errors in these measurements are obtained by block averaging.
Pressure is averaged over four independent runs.


\section{Efficiency analysis}

\subsection{Parameter dependence}		

\begin{figure}
\includegraphics[width=\columnwidth]{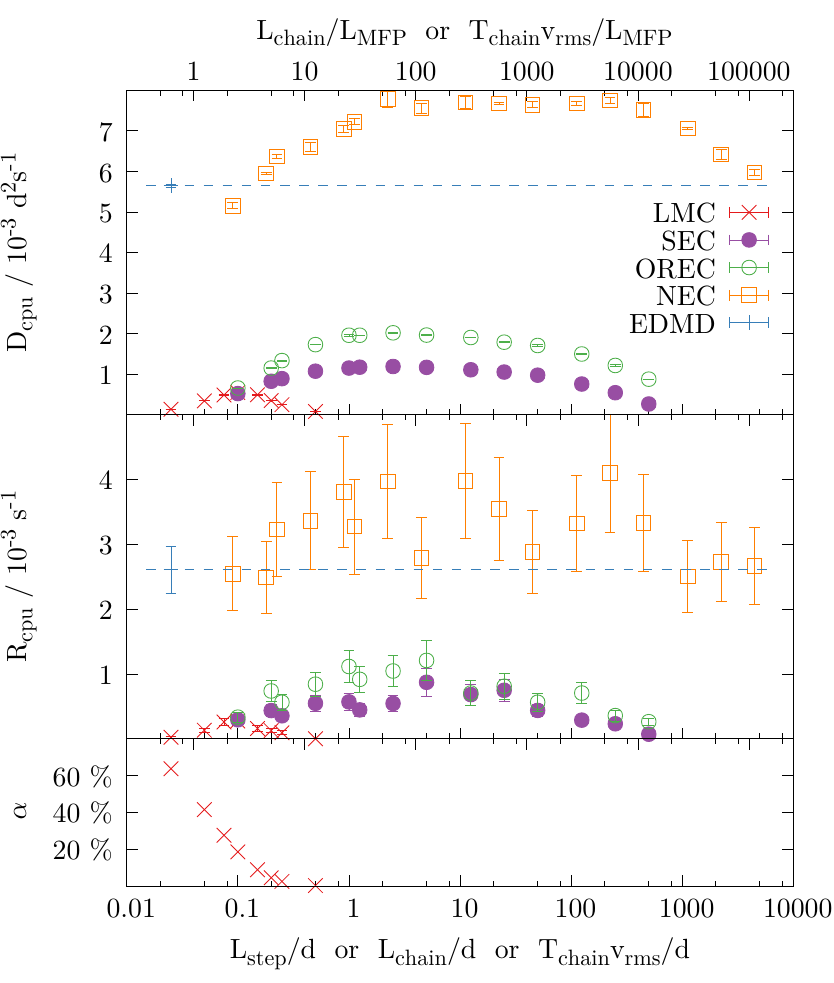}
\caption{
Parameter study of Monte Carlo algorithm efficiency for 16\,384 particles at 54\% volume fraction.
Efficiency is measured by the diffusion coefficient $D_\text{cpu}$ in the fluid before nucleation sets in and the nucleation rate $R_\text{cpu}$ as a function of step size $L_\text{step}$ for LMC, chain length $L_\text{chain}$ for SEC and OREC, and chain duration $T_\text{chain}$ for NEC.
Parameters are given in units of the mean free path $L_\text{MFP}$, the sphere diameter $d$, and the inverse root mean square velocity $v_\text{rms}=\sqrt{\langle v^2\rangle}$ multiplied by a length, respectively.
Data for EDMD is independent of parameters and shown as a dashed line for comparison.
Trial move acceptance probability $\alpha$, the ratio of successful trial moves to attempted trial moves, is given in the bottom plot.
}
\label{fig.parameter}
\end{figure}

The efficiency of Monte Carlo algorithms is affected by the choice of the Monte Carlo parameters step size, chain length, and chain duration.
\cref{fig.parameter} shows the two efficiency measures diffusion coefficient and nucleation rate recorded as a function of these parameters.
A narrow peak for LMC, half an order of magnitude wide, is found for both efficiency measures near $L_\text{step}=3L_\text{MFP}$.
Similar behavior was observed for hard discs in two dimensions.\cite{Bernard2009}
Maximum efficiency occurs near trial move acceptance probability $\alpha=20\%$.
As is often the case, rejected moves are faster to execute than accepted moves due to early returns from the overlap search and the fact that rejected moves do not require an update of the particle position.
The data for SEC and OREC initially follows the curve for LMC and then increases further beyond the LMC peak.
A broad plateau appears for chain length in the range $1<L_\text{chain}/d<100$ in the case of SEC and OREC and for equivalent chain duration in the case of NEC.
OREC performs slightly better and NEC much better than LMC for all chain lengths or chain durations.

We would expect the diffusion coefficient of SEC to match the behavior reported in Ref.~\citenum{Bernard2009} and decrease after the point where a significant part of all particles in the system are translated in the same direction.
In contrast, OREC should not drop even for long chains because the displacement direction keeps getting updated.
We observe in \cref{fig.parameter} that SEC and OREC both decrease in lockstep towards high $L_\text{chain}$.
This means our simulations did not yet reach the point where a sufficient number of particles in the system are moved.
The decrease of $D_\text{cpu}$ must be explained solely due to details of our implementation.
In fact, HOOMD keeps track of particles neighborhoods with an AABB tree data structure.~\cite{Howard2016}
In the current implementation, the AABB tree is only updated but not continuously rebalanced during an event chain.\cite{Anderson2015a}
This means the number of neighbor particle candidates gradually increases as the event chain progresses, which slows down simulations of long event chains.
Efficiency is affected less by overly long chains in NEC.
We do not attempt to improve the use of the AABB tree and select Monte Carlo parameters from the beginning of the plateau.


\subsection{Volume fraction dependence}

\begin{figure} 
\includegraphics[width=\columnwidth]{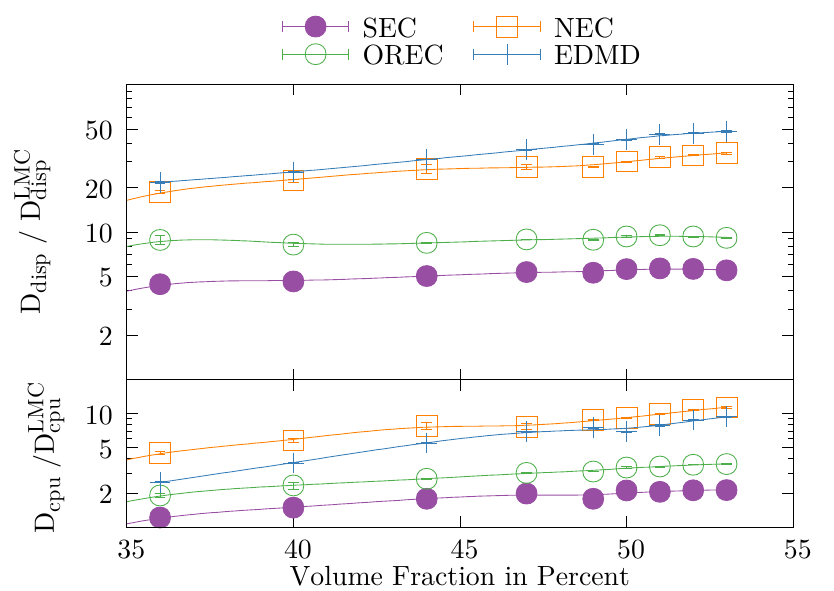}
\caption{
Relative efficiency of hard sphere algorithms measured by the diffusion coefficient as a function of volume fraction in the fluid phase.
All simulations are performed with 16\,384 particles, compared to LMC, and plotted in terms of the number of displacements (top) and CPU time (bottom).
Relative efficiency increases with volume fraction for all algorithms.
For each data point, we performed a parameter study with 30 runs to obtain optimal Monte Carlo parameters for step size and chain length.
Error bars indicate fluctuations.
}
\label{fig.densities}
\end{figure}

Event chain Monte Carlo improves simulation efficiency near the fluid-to-solid transition.\cite{Bernard2009}
Indeed, we observe a performance increase of SEC compared to LMC by a factor of 5 in terms of the number of displacements and 2 in terms of CPU time (\cref{fig.densities}).
The advantage of SEC over LMC keeps growing towards higher volume fraction because particle dynamics becomes more collective.
Still, we do not see quite such a large increase near the critical point as reported in two dimensions.\cite{Bernard2009}
Apparently, collective motion is less important in three dimensions than in two dimensions.

Across all volume fractions, OREC outperforms SEC by a factor of 1.6 but still never reaches the speed of EDMD.
The performance gap between EDMD and all purely stochastic Monte Carlo algorithms grows with volume fraction demonstrating that the Newtonian dynamics of EDMD is superior to stochastic dynamics.
Indeed, the inclusion of Newtonian dynamics into event chain Monte Carlo in the form of NEC outperforms all other algorithms for all densities.


\subsection{System size dependence}

\begin{table*}
	\setlength{\tabcolsep}{3mm}
	\begin{tabular}{l ll ll ll ll}
		\hline
		\hline
		System size
		  &\multicolumn{2}{c}{$N=2^{14}=16\,384$}
		  &\multicolumn{2}{c}{$N=2^{15}=32\,768$}
		  &\multicolumn{2}{c}{$N=2^{16}=65\,536$}
		  &\multicolumn{2}{c}{$N=2^{17}=131\,072$}
		\\
		&$D_{\text{disp}}$& $D_{\text{cpu}}$
		&$D_{\text{disp}}$& $D_{\text{cpu}}$
		&$D_{\text{disp}}$& $D_{\text{cpu}}$
		&$D_{\text{disp}}$& $D_{\text{cpu}}$ \\
		\hline
		opt.\ LMC  &       1.00(1)   &         1.00(1)   
				   &       1.00(1)   &         1.00(3)   
				   &       1.000(6)    &       1.00(1)   
				   &       1.000(2)    &       1.000(2)   \\
		LMC (AP 50\%) &        0.506(7)   &      0.374(8)   
					&       0.506(1)  &       0.384(1)   
					&       0.505(1)   &       0.388(2)   
					&       0.508(1)   &       0.399(1)   \\
		SEC        &        5.134(5)  &       2.12(1)    
					&       5.18(2)  &         2.03(1)    
					&       5.17(2)    &       1.95(3)    
					&       5.213(4)   &       1.91(1)    \\
		OREC       &        8.60(1)   &       3.766(4)   
					&       8.64(2)    &       3.69(1)   
					&       8.644(5)   &       3.742(4)   
					&       8.690(8)   &       3.764(1)   \\
		NEC        &        26.81(4)   &      10.88(2)
					&       27.3(2)    &      10.6(2)
					&       27.23(1)   &      10.30(4)
					&       27.39(1)   &     10.46(4)   \\
		EDMD       &        38.8(3)    &       8.17(5)    
					&       39.4(3)    &       5.47(3)    
					&       39.5(1)    &       5.07(1)    
					&       39.7(3)    &       3.82(3)    
		\\
		\hline
		\hline
	\end{tabular}
	\caption{
		Relative efficiency of hard sphere algorithms measured by the diffusion coefficient for various system sizes and across event chain variants.
		All values are normalized to optimized LMC.
		Simulations were performed at 49\% volume fraction. 
	}
	\label{tab.modes}
\end{table*}

The efficiency of algorithms is compared across system sizes in \cref{tab.modes}.
We choose LMC with step size optimized for maximum diffusion as reference.
The trial move acceptance probability of optimized (opt.)\ LMC is 15\% to 20\%.
In comparison, LMC with acceptance probability 50\% is slower by at least a factor of 2.
As expected, $D_\text{disp}$ is essentially independent of system size.
This makes sense because it only depends on local structure.
In contrast, $D_\text{cpu}$ is affected by system size, most strongly for EDMD.

Because HOOMD-blue and HPMC are highly optimized codes, LMC and event chain Monte Carlo scale well to large numbers of particles.
Our own implementation of EDMD is most efficient for small system size.
Efficiency quickly decreases once system size reaches the CPU cache limit.
We believe our implementation of EDMD can be improved with more effort.
Still, it remains inherently difficult to achieve good scaling to large system sizes for EDMD because collisions can happen anywhere at all times, which makes it practically impossible to utilize CPU cache well.
In contrast, successive collisions in event chains occur in close proximity of another.
For this reason the performance advantage of NEC increases with system size.


\subsection{Melting the FCC crystal} \label{Subsec.melt}

\begin{figure}
\includegraphics[width=\columnwidth]{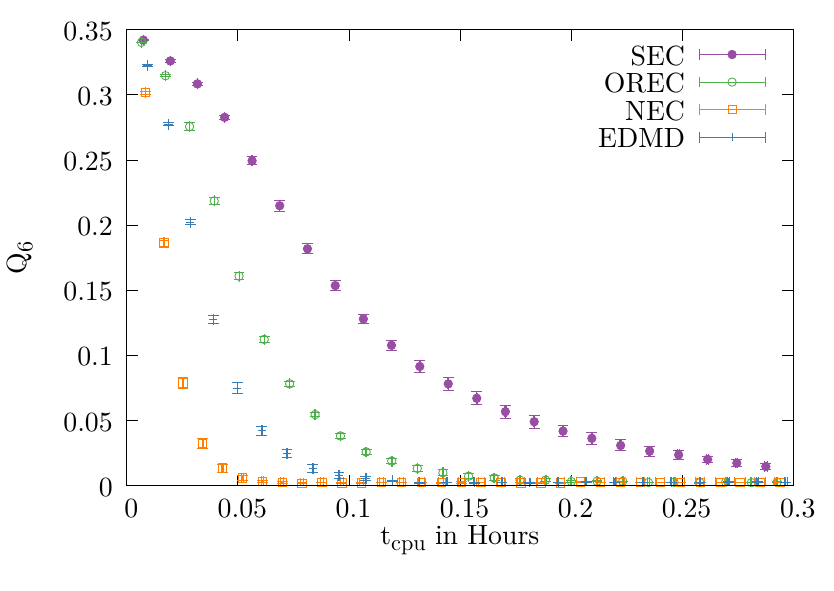}
\caption{
Evolution of the global bond orientational order parameter $Q_6$ during melting of a fcc crystal with 131\,072 hard spheres at 49\% volume fraction.
The fluid has $Q_6$ of near zero.
Data averaged over four trajectories each.
Error bars give the standard error.
}
\label{fig.melting}
\end{figure}

We investigate equilibration by melting in \cref{tab.melting}.
The hard sphere system is initialized in a FCC crystal structure, equilibrated at 55\% volume fraction, and then expanded to 49\%.
All algorithms show a similar melting behavior as quantified by the decrease of the global bond orientational order parameter $Q_6$ (\cref{fig.melting}).
NEC melts within the shortest CPU time, next EDMD, then OREC, and finally SEC.
This sequence is in agreement with measurements of the diffusion coefficient (\cref{tab.modes}).
We observe in \cref{tab.melting} that melting speed is approximately inversely proportional to the diffusion coefficient, which suggests that melting is dominated by diffusion processes.

We remark on an unexpected behavior that we observed in early stages of our tests.
\citet{Isobe2015} propose restricting displacements in SEC in the $+x$, $+y$, and $+z$ directions only.
If we implement such a restriction and align the FCC crystal along the coordinate axes then an artificial speed-up of SEC is observed.
Symmetry axis-aligned event chains melt the crystal faster than event chains with displacement directions not aligned with crystallographic axes and faster than event chains with randomly chosen displacement directions.
To avoid this unphysical dependence of melting on the orientation of the crystal and for better comparison of event chain variants we always choose displacement directions fully randomly in our implementation of SEC.
The use of random displacement directions slightly increases the computational cost for translating particles, which is observed in \cref{tab.speedcomp}.

\begin{table}
	\setlength{\tabcolsep}{3mm}
	\begin{tabular}{l c c c}
		\hline
		\hline
		& $t_\text{cpu}$ to reach & $D_{\text{cpu}}$ & $t_\text{cpu} \times D_\text{cpu}$ \\
		& $Q_6=0.1$ in s             &  in $ 10^{-3} d^2/\text{s}$ & in $d^2$\\
		\hline
		SEC          &     450(20) &  0.76(3) & 0.35(2)  \\
		OREC         &    238(7)  &  1.42(5) & 0.34(2) \\
		NEC          &     85(1)  &  4.7(3)  & 0.40(2)  \\
		EDMD         &    158(7)  &  1.95(7) & 0.31(2)  \\
		\hline
		\hline
	\end{tabular}
	\caption{
		Comparison of melting efficiency of three hard sphere algorithms.
		The table lists the CPU time $t_\text{cpu}$ to melt the majority of an FCC crystal, the diffusion coefficient $D_\text{cpu}$ in the melt, and the product of both.
		Data was averaged over four simulations.
		The error denotes the standard deviation.
		Simulations contain 131\,072 particles at 49\% volume fraction. 
	}
	\label{tab.melting}
\end{table}


\section{Concluding remarks}

\begin{figure}
\includegraphics[width=\columnwidth]{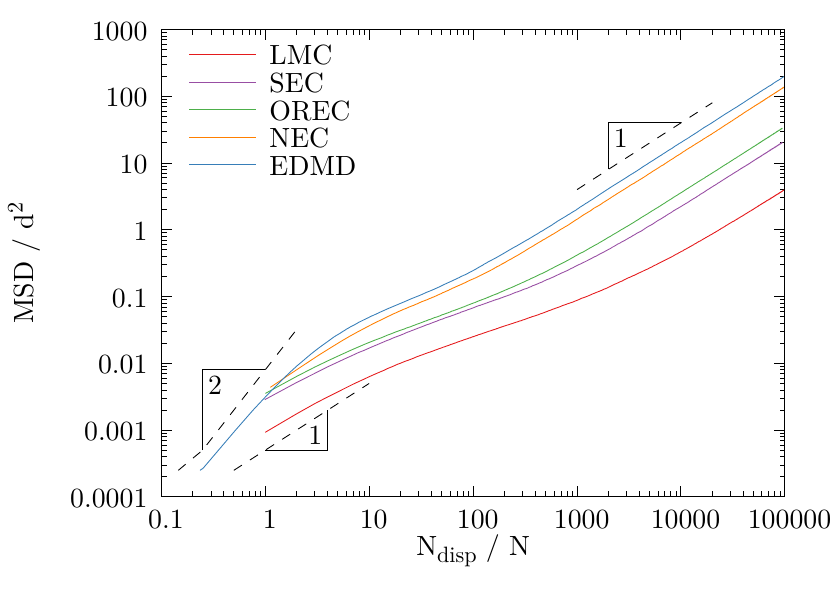}
\caption{
Comparison of mean square displacement (MSD) for various hard sphere algorithms.
Particle motion is characteristically different in EDMD for short times, $N_\text{disp}\lesssim1$, compared to the Monte Carlo algorithms.
Towards large times, $N_\text{disp}\gg1000$, diffusive behavior is restored in all algorithms.
Simulations contain 16\,384 particles at 49\% volume fraction. 
Shown is the average of four trajectories each.
The data spreads about one line width.
}
\label{fig.diffusionscales}
\end{figure}

A central result of this paper is the finding that equilibration trajectories are shorter if particle dynamics more closely mimics Newtonian dynamics.
By bringing the geometry of event chains closer to trajectories generated by molecular dynamics, first in the variant OREC and then in the form of NEC, simulations are sped up significantly.
We believe conservation of momentum in NEC leads to a memory effect that synchronizes the motion of particles in close proximity and allows the system to equilibrate more collectively.

We can understand the effectiveness of particle displacements for advancing the system in configuration space by analyzing mean square displacement in \cref{fig.diffusionscales}.
For short times, i.e.\ for a small number of displacements, $N_\text{disp}\lesssim 1$, ballistic motion in EDMD is responsible for slope 2 scaling while Monte Carlo algorithms have slope $\le 1$.
All algorithms show signs of a slowdown at intermediate times, $10<N_\text{disp}<100$, indicative of caging.
EDMD and NEC gradually increase their advantage over other algorithms and escape caging after the fewest number of displacements.
For long times, $N_\text{disp}\gg 1000$, all algorithms behave indistinguishably and evolve diffusively.
Because displacements in NEC are faster to compute, NEC outperforms EDMD when analyzed in terms of CPU time.

In summary, NEC combines aspects of Newtonian dynamics with event chain Monte Carlo.
NEC is an improvement over past event chain variants and requires little additional implementation effort.
It is currently the algorithms with best CPU performance available for hard spheres.
As all event chain algorithms, NEC scales better than EDMD to large systems.
Future work can generalize NEC to anisotropic hard particles and implement parallelization.


\begin{acknowledgments}
We thank Praveen Bommineni and Sebastian Kapfer for helpful discussions.
This work has been funded by Cluster of Excellence Engineering of Advanced Materials grant EXC 315/2 of the German Research Foundation (DFG).
Support by the Central Institute for Scientific Computing (ZISC), the Interdisciplinary Center for Functional Particle Systems (FPS), and computational resources and support provided by the Erlangen Regional Computing Center (RRZE) are gratefully acknowledged.
\end{acknowledgments}

%

\end{document}